\documentclass[runningheads]{llncs}
\usepackage{graphicx}
\usepackage{makeidx}  %

\usepackage{amsmath,graphicx}
\usepackage{amssymb}
\usepackage{tabulary}
\usepackage{tabularx}
\usepackage{floatflt}
\usepackage{float}
\usepackage{todonotes}
\usepackage{color}

\usepackage{amsmath}
\usepackage{amssymb}

\usepackage{multirow} 
\usepackage{floatflt}
\usepackage{url}
\usepackage{epstopdf}
\usepackage{makecell}
\usepackage{xcolor}
\definecolor{darkgreen}{RGB}{0, 190, 0}

\newcommand{\bfx}{\mathbf{x}}
\newcommand{\bfy}{\mathbf{y}}
\newcommand{\bfb}{\mathbf{b}}

\newcommand{\argmin}{\operatornamewithlimits{argmin}} 
\newcommand{\matr}[1]{\mathbf{#1}}
\newcommand{\vectr}[1]{\mathbf{#1}}
\newcommand*{\tran}{^{\mkern-1.5mu\mathsf{T}}}

\newfloat{myalg}{thp}{lop}
\floatname{myalg}{\textbf{Algorithm}}

\begin{document}
	\title{Deep Variational Networks with Exponential Weighting for Learning Computed Tomography}
	\titlerunning{Deep Variational Networks for CT}
	\author{Valery Vishnevskiy\textsuperscript{1}\and Richard Rau\and Orcun Goksel}  %
	\authorrunning{V. Vishnevskiy et al.}
	\urldef{\mailsa}\path|valeryv@vision.ee.ethz.ch|
	\institute{
	Computer-assisted Applications in Medicine Group, 
		ETH Zurich, Switzerland%
		\\
		\textsuperscript{1}\mailsa
	}
	
	\maketitle              %

	\begin{abstract}

Tomographic image reconstruction is relevant for many medical imaging modalities including X-ray, ultrasound (US) computed tomography (CT) and photoacoustics,  %
for which the access to full angular range tomographic projections might be not available in clinical practice due to physical or time constraints. 
Reconstruction from incomplete data in low signal-to-noise ratio regime is a challenging and ill-posed inverse problem that usually leads to unsatisfactory image quality. 
While informative image priors may be learned using generic deep neural network architectures, the artefacts caused by an ill-conditioned design matrix often have global spatial support and cannot be efficiently filtered out by means of convolutions.
In this paper we propose to learn an inverse mapping in an end-to-end fashion via unrolling optimization iterations of a prototypical reconstruction algorithm.
We herein introduce a network architecture that performs filtering jointly in both sinogram and spatial domains.
To efficiently train such deep network we propose a novel regularization approach based on deep exponential weighting.
Experiments on US and X-ray CT data show that our proposed method is qualitatively and quantitatively superior to conventional non-linear reconstruction methods as well as state-of-the-art deep networks for image reconstruction. 
Fast inference time of the proposed algorithm allows for sophisticated reconstructions in real-time critical settings, demonstrated with US SoS imaging of an \emph{ex vivo} bovine phantom.
\end{abstract}

\section{Introduction}
Tomographic image reconstruction with sparse or limited angular (LA) data arises in a number of applications including image guided interventions~\cite{siewerdsen2007multimode}, photoacoustics~\cite{lin2018evaluation_photoac}, and US speed-of-sound (SoS) imaging~\cite{sanabria2016hand,cheng2019deep}.
Such underdetermined problems usually require suitable problem-specific regularization for meaningful reconstructions, e.g.\ free from streaking artefacts.
Setting regularization parameters manually can be cumbersome and often generalizes poorly. 
Using learning-based methods as %
in~\cite{zheng2018pwls} can greatly improve reconstruction accuracy and account for non-Gaussian noise models.
Unfortunately the method in~\cite{zheng2018pwls} is based on patch-based clustering leading to very slow reconstruction.
Straightforward application of computationally efficient convolutional network directly to measurements might be tempting, but is unjustified, because sinogram values have global spatial dependence on image intensities.
In practice, such generic networks are not likely to generalize well for LA-CT problems~\cite{maier2019gentle}.
Many deep-learning inspired methods employ artificial neural networks to learn filtering or weighting~\cite{wurfl2018deep,schwab2019learned} of the input sinograms prior to the backprojection step, after which the result might be postprocessed by another network~\cite{hammernik2017deep}.
Unfortunately such sinogram preprocessing requires problem-specific weighting schemes, which would constrain applicable acquisition geometries.
Variational Networks (VN) employ adjoint of projection operator to learn convolutional filters in \emph{spatial} domain.
For compressed sensing in MRI, a landmark VN architecture was introduced by Hammernik et al.~\cite{hammernik2018learning}, which in practice relies on unitarity of Fourier transform.
This was addressed in~\cite{vishnevskiy2018image} with more sophisticated unrolled iterations that improved USCT reconstructions and allowed \emph{detection} of coarse blob-looking inclusions.

In order to allow for accurate image reconstruction with ill-conditioned spatial encoding operators, in this paper we extend VN architecture by introducing sinogram filters that are learned as preconditioners, inspired by filtered backprojection.
We also propose an efficient network regularization scheme that allows stable training inspired by Landweber iterations~\cite{landweber1951iteration} and deep supervision~\cite{liu2016learning}.

\section{Methods}
Tomographic reconstruction problem involves estimating image intensities $x$ from set of measurements (sinogram) $b_i$ (e.g., time-of-flight or ray attenuation) that are modelled, e.g., as line integrals 
$b_i=\int_{\text{ray}_i}x(\mathbf{r})ds$.
Given a set of measurements $\bfb\in\mathbb{R}^M$, algebraic reconstruction methods solve for discretized spatial encoding equations in a maximum-a-posteriori (MAP) sense, i.e.:
\begin{equation}
\label{eq:art_recon}
    \hat{\bfx}(\bfb; \lambda, p) = \argmin_\bfx \|\matr{L}\bfx-\bfb\|_p^p +\lambda\mathcal{R}(\bfx),
\end{equation}
where $\bfx\in\mathbb{R}^{n_1n_2}$ is $n_1$$\times$$n_2$ image and
$\matr{L}\in\mathbb{R}^{M\times n_1n_2}$ is a sparse, ray-discretization matrix that depends on the acquisition geometry. 
The norm $p$ determines data inconsistency penalty, e.g.\ $p$$=$$2$ assumes Gaussian acquisition noise while $p$$=$$1$ assumes Laplace noise, the latter of which is often considered to be more robust.
Non-negative weight $\lambda$ controls the influence of regularizer. 
Similarly to~\cite{knoll2011second}, we herein consider total variation 
$\mathcal{R}_\text{TV}(\bfx)$=$\|\boldsymbol{\nabla}\bfx\|_1$
and total generalized variation 
$\mathcal{R}_\text{TGV}(\bfx)$=$\min_\vectr{u}\|\nabla\bfx-\vectr{u}\|_1+2\|\boldsymbol{\mathcal{E}}\vectr{u}\|_1$ regularizers.
Here $\boldsymbol{\nabla}$ denotes first-order forward finite derivative matrix and $\boldsymbol{\mathcal{E}}$ is the symmetrized vector field derivative operator.
Both TV and TGV regularizers yield convex optimization problems for $p=\{1,2\}$, which we hereafter refer as L$p$TV and L$p$TGV.

\begin{figure}[t!]
	\includegraphics[width=\textwidth]{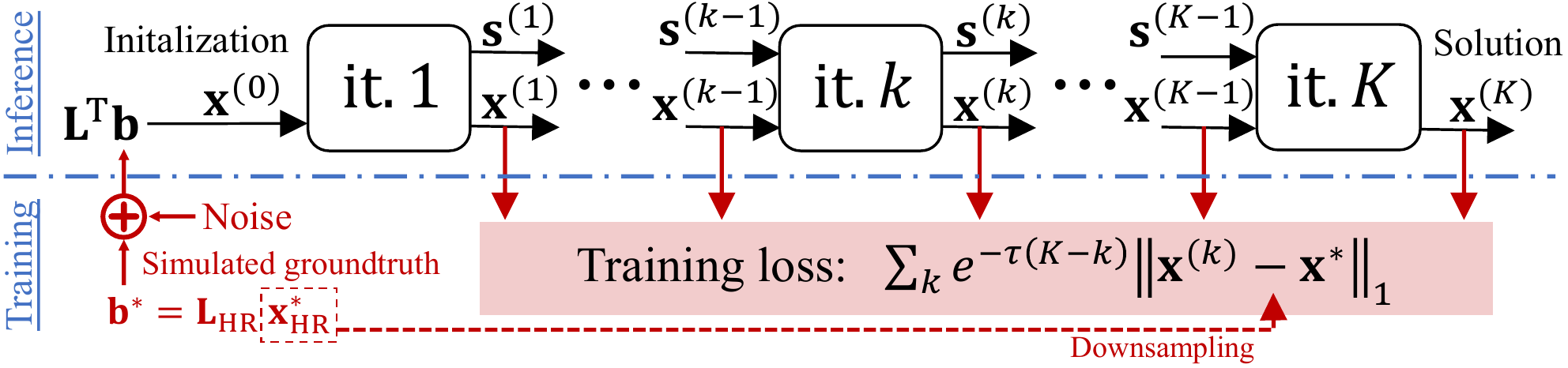}
	\vspace{-5mm}
	\caption{
	    Structure of variational network and its training strategy. 
	}
	\label{fig:scheme}
\end{figure}
A Variational Network can be seen as a sequence of $K$ unrolled iterations of a numerical optimization scheme, inspired by a prototypical objective as in~(\ref{eq:art_recon}).
For additional learning capacity, these iterations are further relaxed, e.g.\ by adding variable filters and activations that can be tuned during training.
As illustrated in Fig.~\ref{fig:scheme}, we initialize VN inference by $\matr{L}\tran\bfb$.
Then, at each network layer (iteration) $k$, a \emph{gradient} term $\vectr{g}^{(k)}$ is accumulated via running average governed by momentum coefficient $\alpha^{(k)}$ and used to
update current image estimate $\bfx^{(k)}$ as follows:
\begin{equation}
\begin{aligned}
    \label{eq:grad_term}
    &\vectr{g}^{(k)}\leftarrow
    \left(\matr{P}^{(k)}\matr{L}\matr{Q}^{(k)}\right)\tran
    \matr{W}_d^{(k)}\varphi_d^{(k)}\left\{
        \matr{W}_d^{(k)}\left(\matr{P}^{(k)}\matr{L}\matr{Q}^{(k)}\bfx^{(k-1)}-\bfb\right)
    \right\} + \\
    &\quad\quad\quad\;\left(\matr{D}^{(k)}\right)\tran\matr{W}_r^{(k)}\varphi_r^{(k)}\left\{
        \matr{W}_r^{(k)}\matr{D}^{(k)}\bfx^{(k-1)}\right\},\\
    &\vectr{s}^{(k)}\leftarrow\alpha^{(k)}\vectr{s}^{(k-1)} + \vectr{g}^{(k)}, \quad\quad
    \vectr{x}^{(k)}\leftarrow\vectr{x}^{(k-1)}-\vectr{s}^{(k)}.
\end{aligned}
\end{equation}
We define the gradient term $\vectr{g}^{(k)}$ with the following tunable operations that all allow backpropagation of gradients:
(i)~multiplication with diagonal \emph{preconditioner} $\matr{W}_d^{(k)}$ and spatial regularization \emph{weighting} $\matr{W}_r^{(k)}$; 
(ii)~convolution with left ($\matr{P}^{(k)}$) and right ($\matr{Q}^{(k)}$) preconditioners, and several (herein, $n_\mathrm{f}$$=$$50$) regularization filters $\matr{D}^{(k)}$;
(iii)~nonlinear data ($\varphi_d^{(k)}$) and regularization ($\varphi_r^{(k)}$) activation functions that are parametrized via linear interpolation on a regular grid, herein, of size $n_\mathrm{g}$$=$$35$; i.e.\
$\varphi\{t\}=(1 - t + \lfloor t\rfloor)\phi_{\lfloor t\rfloor}+(t - \lfloor t\rfloor)\phi_{\lfloor t\rfloor+1}.$
To avoid bilinear ambiguities, every $n_\mathrm{k}$$\times$$n_\mathrm{k}$ (herein, $n_\mathrm{k}$$=$$7$) convolution $\matr{D},\matr{Q},\matr{P}$ with kernel $\vectr{d}$ is reparametrized to be zero-centered unit-norm, i.e.\ 
$\vectr{d}=n_\mathrm{k}(\vectr{d}'-\text{mean}(\vectr{d'}))/\text{std}(\vectr{d'})$,
while diagonal terms are also ensured to be nonnegative and bounded via sigmoid: $\matr{W}=\text{diag}(\sigma(w_i))$.
Stochastic minimization of the \emph{exponentially weighted} $\ell_1$ reconstruction loss is then conducted to tune parameter set 
$\Theta$=$\{\matr{P}^{(k)}\!, \matr{Q}^{(k)}\!, \matr{D}^{(k)}\!, \matr{W}_r^{(k)}\!, \matr{W}_f^{(k)}\!, \boldsymbol{\phi}_r^{(k)}\!, \boldsymbol{\phi}_d^{(k)}\!, \alpha^{(k)}\!\}$:
\begin{equation}
\label{eq:exp_loss}
    \min_\Theta \mathop{\mathbb{E}}_{ \{\bfb, \bfx^\star\} \in \mathcal{T}}
    \sum_{k=1}^K
		e^{-\tau(K-k)}\,\| \bfx^{(k)}(\bfb; \,\Theta) - \bfx^\star\|_1,
\end{equation}
on the training set $\mathcal{T}$.
Here $\tau$$\geq$$0$ controls the regularization of the network: at $\tau$$=$$0$, the reconstruction on all layers is weighted equally, therefore all network parameters have low variance of gradients, which allows for stable training.
For $\tau$$\rightarrow+$$\infty$, only the last network output $\bfx^{(K)}$ is used for training, which allows the network to be tuned accurately for the final objective.
We accordingly increase $\tau$ during the training procedure to gradually relax constraints on the network.
Intuitively, such regularization encourages VN to provide reconstruction as early as possible, which is inspired by \emph{early stopping} --- a common image reconstruction strategy that allows to avoid degenerate solutions and can be shown to be equivalent to Tikhonov regularization in certain cases~\cite{landweber1951iteration}.

\vspace{1ex}\noindent {\bf Training. }
We employ a $K$=10 layer VN and perform 5$\cdot$$10^4$ iterations of Adam algorithm (learning rate $10^{-3}$, $\beta_1$$=$$0.85$, $\beta_1$$=$$0.98$, batch size of 10) for training, during which we continually adjust $\tau=j$$\cdot$$10^{-3}$ with $j$ being the iteration number.
To avoid overfitting to the discretization scheme employed in the simulation, we use higher resolution images $\bfx_\text{HR}$ to compute line integrals defined by $\matr{L}_\text{HR}$ while training for these to be reconstructed in the desired resolution defined by the discretization of $\matr{L}$.
The network was implemented using Tensorflow framework, where multiplication with the design matrix $\matr{L}$ was carried out as a generic sparse matrix-vector multiplication.
For comparison with iterative reconstruction, we employ ADMM algorithm~\cite{boyd2011distributed} to solve~(\ref{eq:art_recon}) by approximating the regularized inversion of $\matr{L}$ with 5 iterations of LSQR solver~\cite{paige1982lsqr}.
For each experimental scenario, an optimal value of $\lambda$ was tuned on a single test image.

\begin{figure}[t]
\label{fig:abl}
\includegraphics[width=\textwidth]{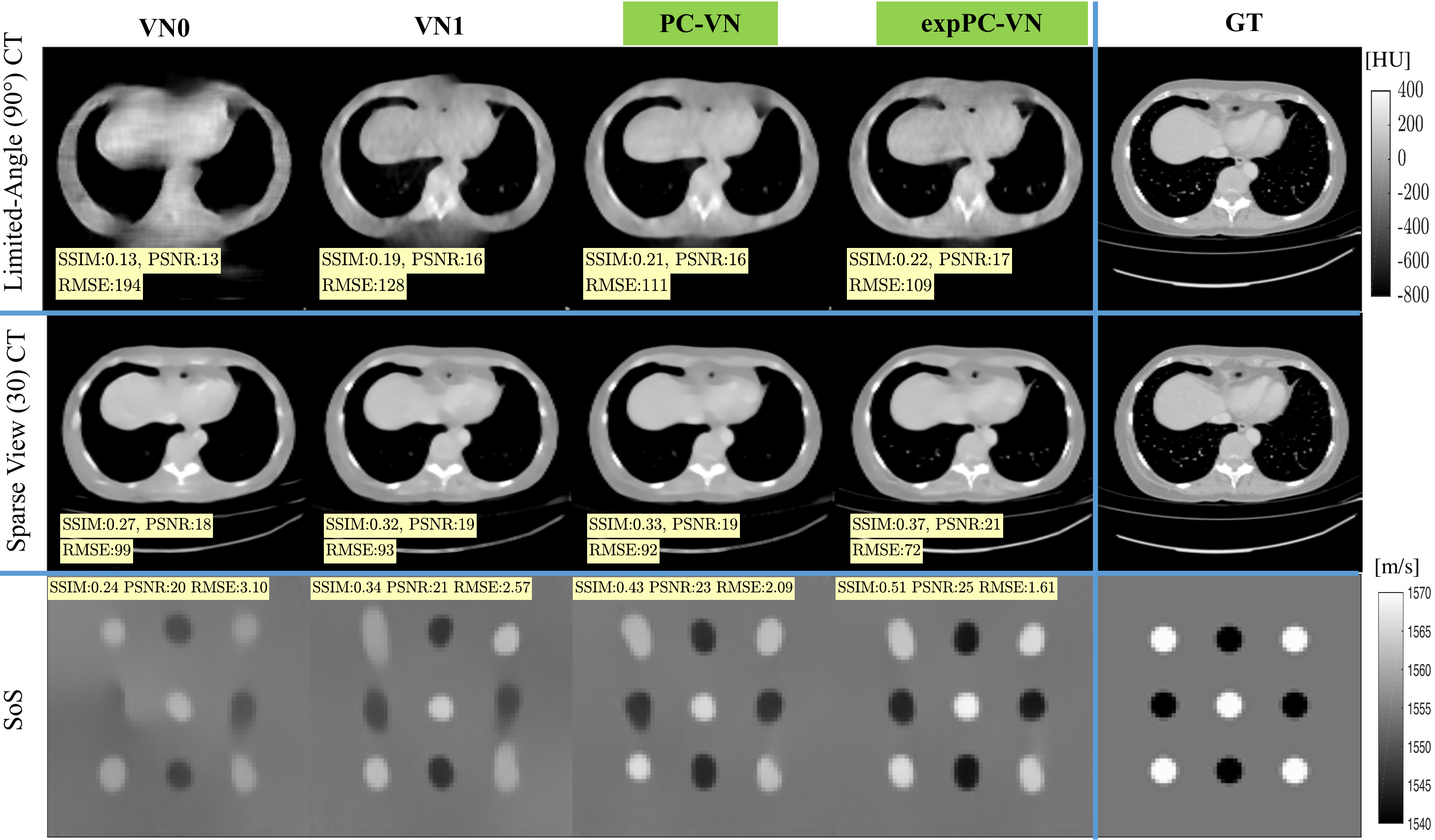}
\caption{
Performance comparison of VN architectures on X-ray CT and SoS simulations. 
Proposed architectures are highlighted in green.
}
\end{figure} 

\section{Experiments and Results}
{\bf X-ray CT dataset. }
We used 3DIRCADb dataset~\cite{soler20103d} that consists of 3080 axial CT scans from 22 patients with inplane resolution varying from 0.56$^2$ to 0.961$^2$\,mm$^2$ and slice thickness of 1.6 to 4\,mm.
The forward simulation was conducted on original 512$\times$512 images, which then were downsampled to 256$\times$256 grid, yielding the ground truth (GT). 
We simulated parallel beam acquisition geometry for limiter-angle (LA) and sparse-view (SV) scenarios.
In the LA scenario we simulated angular ranges of 120$^\circ$, 90$^\circ$ and 60$^\circ$, where projections 
were acquired in 1$^\circ$ increments.
For the SV scenario we simulated 180$^\circ$ range with 60, 30, and 15 uniformly-acquired projections.
To simulate realistic acquisition noise, we follow~\cite{zheng2018pwls} and employ Poisson+Gaussian model, i.e.\ \mbox{$b=\log\big|\text{Poisson}(I_0\exp(-b^\star))$$+$$\text{Gauss}(0, \sigma_E)\big|$}\,, with $I_0$=2$\cdot10^4$ and $\sigma_E$=8$\cdot\text{{Unif}}(0, 1)$ to allow variable SNR.
We used 20 patient scans for training and two for testing.

\vspace{1ex}\noindent {\bf US Speed-of-Sound Tomography.}
We follow~\cite{vishnevskiy2018image} to simulate reflector-based USCT reconstruction with a 128-element transducer and a square imaging field-of-view. 
Synthetic inclusion masks were generated at 256$\times$256 resolution as levelsets of random, spatially-smooth functions. 
The inclusion SoS values were then randomly sampled from $[1350, 1650]$\,m/s. 
Acquisition noise was modelled as Gaussian with $\sigma_N$=$2\cdot$10$^{-8}$ and the reconstruction was conducted on a coarser 64$\times$64 grid to avoid overfitting to the discretization.
The training set contains 15000 random synthetic inclusions, while the test set includes 13 geometric primitives consisting of oval and polygonal inclusions.

	\begin{table}[t]
		\begin{center}
			\caption{
			Mean reconstruction RMSE computed on corresponding training sets with standard deviations indicated in parentheses. Proposed methods are highlighted.
			}
			\label{tbl:res_table}
			\scriptsize
			\begin{tabulary}{\linewidth}{l|C|C|C|C|C|C|C}
				\hline 
				\textbf{Dataset} & 
				{\textbf{L2TV}}& 
				{\textbf{L1TV}}&
				{\textbf{L2TGV}}& 
				{\textbf{VN0}}& 
				{\textbf{VN1}}&
				{ \colorbox{green!40}{\textbf{PC-VN}}}&
				{ \colorbox{green!40}{\textbf{expPC-VN}}}\\
				~ & RMSE& RMSE& RMSE& RMSE& RMSE& RMSE& RMSE \\
				\hline
				\makecell[l]{{LA-CT} 90$^\circ$} 
				&216\,(9)&238\,(16)& 205\,(6)&210\,(9)&128\,(15)&119\,(19)&{\bf 103\,(8)}\\
				\makecell[l]{{SV-CT} 30} 
				&93\,(5)&91\,(4)& 98\,(6)&97\,(4)&87\,(8)&88\,(9)& {\bf 65 (5)}\\
				\makecell[l]{SoS USCT} 
				&1.48\,(0.44)&1.62\,(0.48)& 1.89\,(0.35) & 1.8\,(0.33) & 1.35\,(0.43)&1.19\,(0.41)&{\bf 1.0\,(0.35)}\\
				\hline
			\end{tabulary}
		\end{center}
	\end{table}

\begin{figure}[!h!]
\label{fig:cmp}
\includegraphics[width=0.97\textwidth]{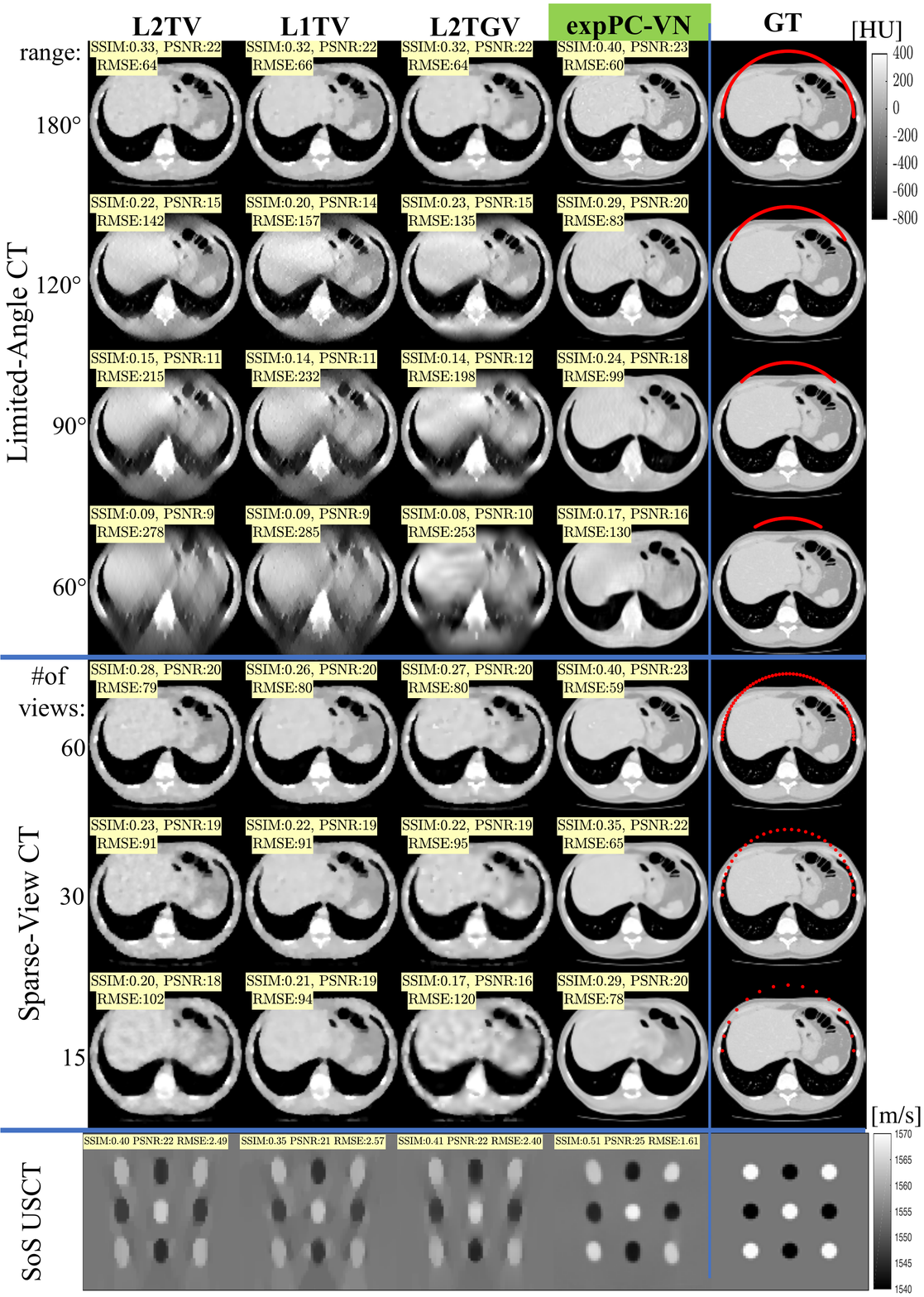}
\caption{
Reconstruction results for X-ray CT and USCT acquisitions.
For sparse view and limited angle experiments, acquired angular positions of projections are depicted in the GT column.
}
\end{figure}

\vspace{1ex}\noindent {\bf Evaluation. } For quantitative comparison of reconstruction and ground truth, we calculated structural similarity index measurement (SSIM)~\cite{wang2004image}, Root Mean Square Error (RMSE), and peak signal-to-noise ratio (PSNR) as follows:
\begin{equation}
    \text{RMSE}(\bfx, \bfy)=\sqrt{\frac{\|\bfx-\bfy\|_2^2}{N}},\qquad
    \text{PSNR}(\bfx, \bfy)=10\log_{10}\frac{R^2N}{\|\bfx-\bfy\|_2^2},
\end{equation}
where $R$ is the dynamic range of the ground truth image.
The corresponding values are reported in Hounsfield units (HU) and m/s, for X-ray CT and USCT SoS experiments accordingly.
We denote the architectures employed in~\cite{hammernik2018learning} and~\cite{vishnevskiy2018image} as VN0 and VN1, respectively.
As seen in Fig.~\ref{fig:abl} the proposed preconditioned network (PC-VN) with sinogram convolutions improves reconstruction accuracy and quality for all X-ray CT acquisition scenarios and USCT as suggested by RMSE and SSIM values. 
As reported in Tab.~\ref{tbl:res_table}, training the proposed network using exponentially weighted loss (expPC-VN) defined in Eq.~(\ref{eq:exp_loss}) further improves reconstruction quality and reduces the variance of error, which can be explained by the introduced regularization effect.
Fig.~\ref{fig:cmp} shows that expPC-VN outperforms iterative methods both in terms of accuracy and image quality.
Namely, compared to nonlinear reconstruction methods, we observe improvements of RMSE by 49\% in the CT-LA-60$^\circ$ scenario, and increase of SSIM by 38\% in the CT-SV-15 experiment.
Quantitative results from Tab.~\ref{tbl:res_table} and Fig.~\ref{fig:cmp} show that proposed reconstruction method outperforms all considered iterative and deep learning -based approaches.

\begin{figure}[t]
\label{fig:fintest}
{\hspace{25mm}(a)\hspace{55mm}(b)}\\
\includegraphics[width=\textwidth]{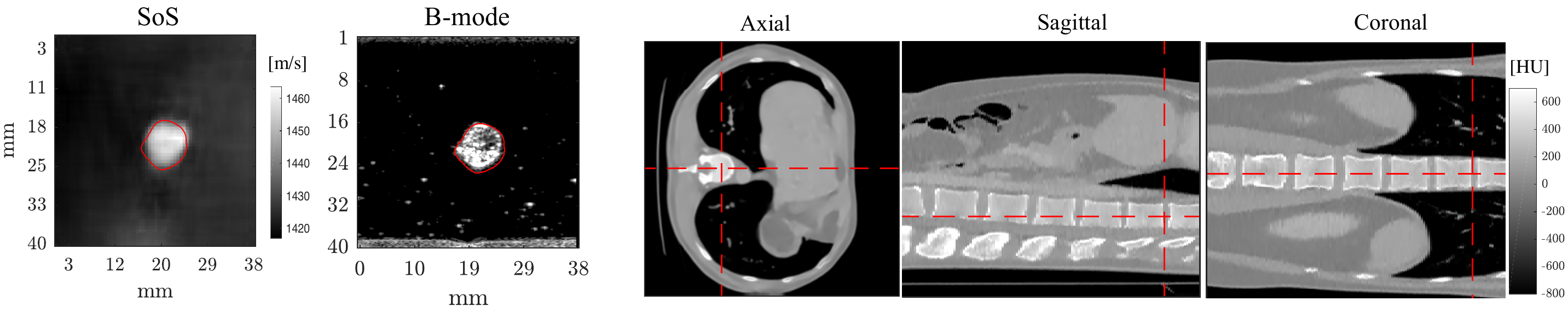}
\caption{
(a) expPC-VN SoS reconstruction of \emph{ex vivo} phantom and corresponding B-mode image. 
Red contour shows inclusion segmentation from the B-mode image.
(b) Axial, sagittal and coronal view of slice-wise reconstruction of the proposed expPC-VN reconstruction network. Corresponding slice positions are indicated with red dashed line.
}
\end{figure}

In Fig.~\ref{fig:fintest}\,(a), we present a USCT SoS reconstruction from \emph{ex vivo} bovine skeletal muscle tissue embedded in a gelatin phantom. 
Compared to the conventional B-Mode image, we could accurately identify inclusion location and provide quantitative estimates of local tissue SoS.
Reconstruction of a single image with VN takes 0.03\,s on NVIDIA Titan Xp GPU  and 1-4\,min with iterative methods on a 6-core 3.7\,GHz Intel CPU.
In order to demonstrate potential 3D imaging applications of our method,
we also conducted X-ray CT reconstruction of SV-60 acquisition scenario with test images rotated by 90$^\circ$ in the axial plane, and show a cross-sectional views from the reconstructed 3D volume in Fig.~\ref{fig:fintest}\,(b).
We observe high spatial coherence and contrast in coronal and sagittal planes which asserts high generalization ability of the proposed expPC-VN method.

\section{Conclusions}
In this paper we have presented a network architecture for preconditioned reconstruction and a regularization scheme for its efficient training via exponential weighting.
The proposed network has been shown to outperform conventional algebraic and learning-based reconstruction methods in terms of accuracy and image quality for various challenging X-ray CT and SoS USCT scenarios. 
Such effectiveness and versatility of our approach may suggest its potential for solving other intriguing optimization and inverse problems also outside of the image reconstruction field.

	\bibliographystyle{splncs04}
	\bibliography{refs}
	
\end{document}